\documentclass[prb,aps,twocolumn,superscriptaddress,10pt,amsmath,amssymb,nofootinbib,floatfix,longbibliography]{revtex4-1}
\pdfoutput=1
\usepackage{epsfig}
\usepackage{amsmath}
\usepackage{amssymb}
\usepackage{graphicx}
\usepackage{enumitem}
\usepackage{longtable}
\usepackage{dcolumn}

\newcommand*{\bb}[1]{\mathbf{#1}}

\newcolumntype{L}{D{.}{.}{2,2}}

\makeatletter 

\begin{document}

\title{On the possibility that PbZrO$_3$ not be antiferroelectric}

\author{Hugo Aramberri}
\affiliation{Materials Research and Technology
  Department, Luxembourg Institute of Science and Technology, 5 avenue
  des Hauts-Fourneaux, L-4362 Esch/Alzette, Luxembourg}

\author{Claudio Cazorla}
\affiliation{Departament de F\'{i}sica, Universitat
Polit\`{e}cnica de Catalunya, Campus Nord B4-B5, Barcelona 08034, Spain}

\author{Massimiliano Stengel}
\affiliation{Institut de Ci\`{e}ncia de Materials
de Barcelona (ICMAB-CSIC), Campus UAB, 08193 Bellaterra, Spain}
\affiliation{ICREA-Instituci\'{o} Catalana de Recerca i Estudis Avan\c{c}ats, 08010 Barcelona, Spain}

\author{Jorge \'I\~niguez}
\affiliation{Materials Research and
  Technology Department, Luxembourg Institute of Science and
  Technology, 5 avenue des Hauts-Fourneaux, L-4362 Esch/Alzette,
  Luxembourg} 
\affiliation{Department of Physics and Materials
  Science, University of Luxembourg, 41 Rue du Brill, Belvaux L-4422,
  Luxembourg}

\date{\today}
\begin{abstract}
Lead zirconate (PbZrO$_{3}$) is considered the prototypical antiferroelectric material with an antipolar ground state. Yet, several experimental and theoretical works hint at a partially polar behaviour in this compound, indicating that the polarization may not be completely compensated. In this work we propose a simple ferrielectric structure for lead zirconate. First-principles calculations reveal this state to be more stable than the commonly accepted antiferroelectric phase at low temperatures, possibly up to room temperature, suggesting that PbZrO$_{3}$ may not be antiferroelectric at ambient conditions. We discuss the implications of our discovery, how it can be reconciled with experimental observations and how the ferrielectric phase could be obtained in practice.
\end{abstract}
\maketitle

\vspace{5mm}{\bf Introduction}

Lead zirconate is often considered the prototypical antiferroelectric material and was the first compound identified as such~\cite{Sawaguchi1951}. Its antipolar structure has been a matter of investigation for several decades, both experimentally~\cite{Sawaguchi1951,jona1955,jona1957,fujishita1982crystal,glazer1993structure,teslic1998atomic,hlinka14} and theoretically~\cite{singh1995structure,waghmare1997,singh1997,kagimura2008first,tagantsev2013origin,reyes2013antiferroelectricity,iniguez2014first,baker2021re}. The antiferroelectric state of PbZrO$_3$, which displays oxygen octahedra rotations and antiparallel displacements of the lead atoms, is described by a 40 atom unit cell with space group symmetry $Pbam$ and is currently considered to be the ground state of this compound (we will refer to this state as `AFE$_{40}$' in the following). A recent work~\cite{baker2021re} predicted from first-principles a new ground state for PbZrO$_3$, characterized by a small cell-doubling distortion of the AFE$_{40}$ state but retaining the same antipolar pattern (`AFE$_{80}$' in the following).

Previous theoretical works on PbZrO$_3$ have reported polar polymorphs~\cite{kagimura2008first,reyes2013antiferroelectricity,iniguez2014first} lying at relatively low energies, and antipolar polymorphs with longer period modulations than the AFE$_{40}$ phase have also been found. Other theoretical investigations indicate that PbZrO$_3$ would present an incommensurate phase~\cite{tagantsev2013origin}. Incommensurate phases have indeed been stabilized in PbZrO$_3$ with small pressures~\cite{burkovsky2017critical}, and in several PbZrO$_3$-based materials by doping~\cite{ma2019uncompensated,fu2020unveiling}. Also, some experimental works report a combination of ferroelectric and antiferroelectric behaviours in this compound~\cite{jona1955,PZO-Raman,roleder00}, and even an intermediate ferroelectric phase has been reported in a narrow temperature range between the cubic and antiferroelectric states~\cite{tennery1965study,tanaka1982,tagantsev2013origin}. Moreover, polar antiphase boundaries have been often reported and studied in PbZrO$_3$~\cite{jona1955,tanaka1982,wei2014ferroelectric,wei2015preferential,wei2015polarity,vaideeswaran2015search,puchberger2016domain,schranz2020contributions}. These results suggest that the AFE$_{40}$ phase (or, for that sake, the recently proposed AFE$_{80}$), might not actually be the ground state of this compound. Instead, a state with a different-period dipole ordering, maybe featuring an only partial dipole compensation (ferrielectric), might be the preferred low temperature structure of PbZrO$_3$.

In this work we introduce a simple ferrielectric phase for PbZrO$_3$ that is stable. We use density functional theory to compare the stability of this polymorph with the commonly accepted antipolar AFE$_{40}$ ground state, the closely related AFE$_{80}$ antipolar phase, and one of the known low energy polar polymorphs with rhombohedral symmetry. The free energy dependence of these phases with temperature reveals that the commonly accepted ground state is the most stable polymorph at the usual growth temperatures, but upon cooling the ferrielectric phase becomes more stable. Our work thus indicates that the model perovskite antiferroelectric might actually be ferrielectric at low temperatures, maybe even at ambient conditions.

\vspace{5mm}{\bf Results}

\begin{figure}
 \includegraphics[scale=0.055]{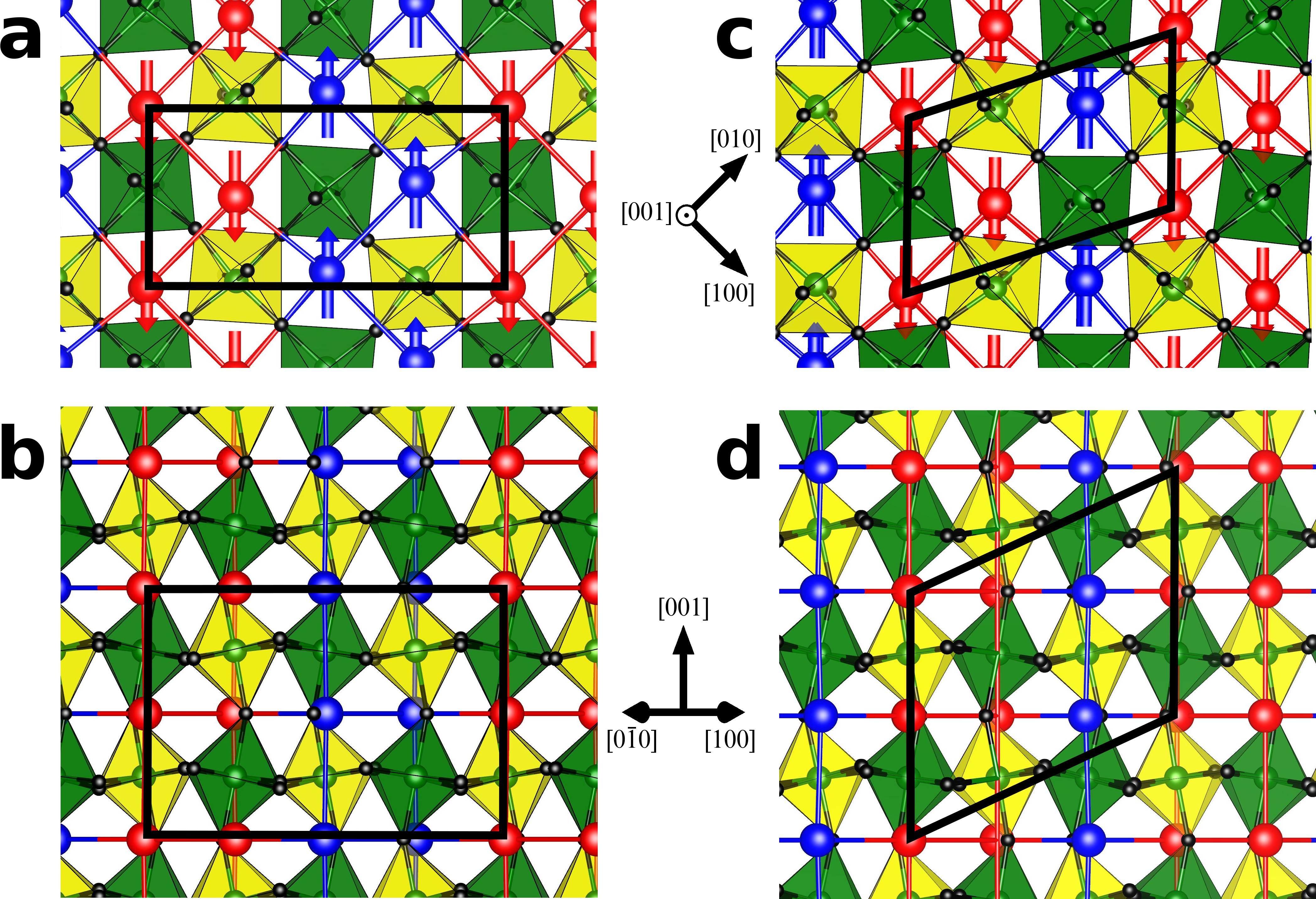}
	\caption{{\bf Crystal structure of the AFE$_{40}$ and FiE phases of PbZrO$_3$.} Top ({\bf a},{\bf c}) and side ({\bf b},{\bf d}) views of the  AFE$_{40}$ (left panels) and FiE states (right panels) are displayed. The primitive cells are marked with a black line. The Zr and O atoms are shown in green and black, respectively. Pb atoms are coloured in red and blue for positive and negative displacements, respectively. The oxygen octahedra are coloured in green and yellow alternatively (according to their displacements given by the $R$-point instability) as a guide to the eye.}
	\label{fig_crystal}
\end{figure}

\begin{figure}
 \includegraphics[scale=0.06]{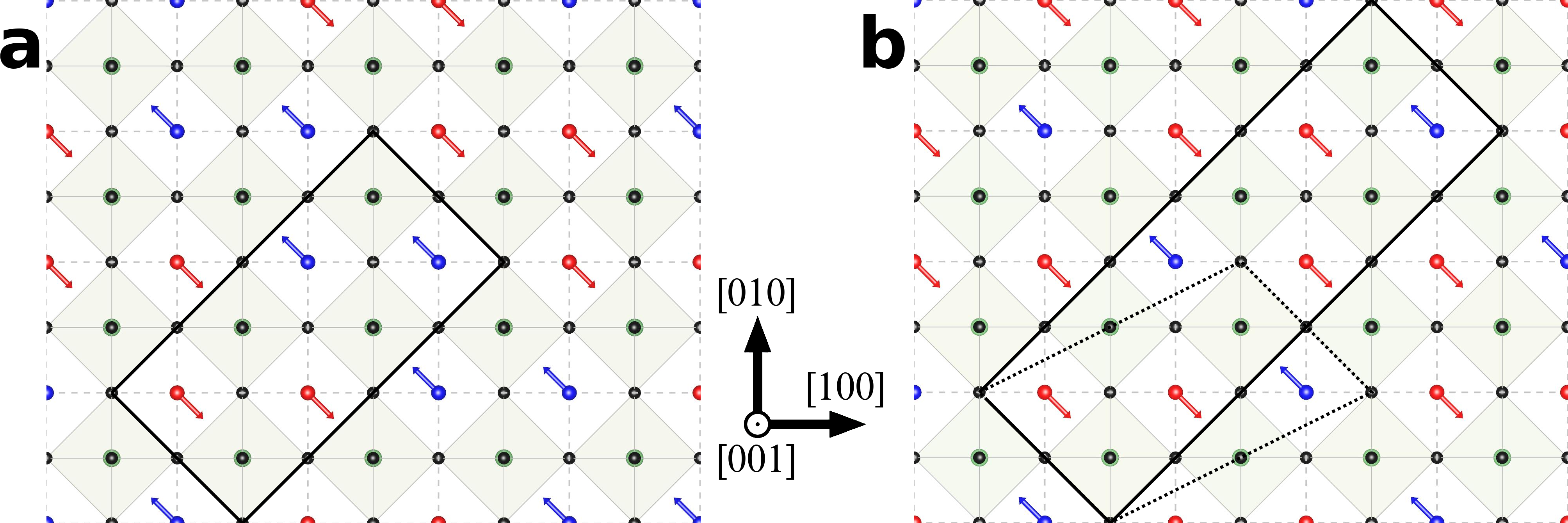}
	\caption{{\bf Schematic Pb displacements in the AFE$_{40}$ and FiE phases of PbZrO$_{3}$.} Top view of the in-plane displacements of the lead cations in the AFE$_{40}$ phase ({\bf a}), and in the FiE phase ({\bf b}). Unit cells marked in black; lead cations marked in red and blue. For the FiE phase a smaller primitive cell (of only 30 atoms) exists, for which the in-plane projection is shown with a dotted line in  {\bf b} (see also Fig.~\ref{fig_crystal}). As explained in the text, in {\bf b} the two red cations in the primitive cell move by $+$0.16~\AA\ and $+$0.23~\AA, respectively, while the blue cation moves by $-$0.21~\AA.}
	\label{fig_crystal_schem}
\end{figure}

{\bf Basic DFT results.} We use first-principles simulations based on density functional theory (see Methods) to study the stability of some relevant PbZrO$_{3}$ phases. 
The optimized crystal structure of the AFE$_{40}$ phase is shown in Figs.~\ref{fig_crystal}{\bf a} and~\ref{fig_crystal}{\bf b}, and displays antipolar displacements of the Pb atoms by 0.19~\AA\ along the $[1\bar{1}0]$ pseudocubic axis. (We employ the pseudocubic reference throughout.) Repeating the analysis in Ref.~\onlinecite{iniguez2014first}, we use standard crystallographic tools (see Methods) to identify the symmetry-adapted distortions connecting it to the cubic perovskite parent phase. The leading $R$-point instability of the cubic phase (see Suppl. Fig. 1) leads to the condensation of an $R_{4}^{+}$ mode associated to antiphase oxygen octahedra rotations with an $a^-a^-c^0$ tilt pattern in Glazer's notation~\cite{glazer1972classification}. This mode accounts for 59.8\% of the total distortion. The second largest distortion is a $\Sigma_{2}$ mode at $\bb{q}=(\frac{1}{4},\frac{1}{4},0)$ which involves antipolar displacements of the Pb cations by 0.26~\AA\ along $[1\bar{1}0]$, as well as a deformation of the oxygen octahedra, and accounts for 36.3\% of the total distortion. The atomic dipoles resulting from this mode follow the \textit{up-up-down-down} pattern that is the best-known feature of the AFE$_{40}$ state (see Fig.~\ref{fig_crystal_schem}{\bf a}). The remaining modes account for 3.6\% of the total distortion. Note that we use symmetry labels corresponding to the crystallographic setting with the origin at the Zr atom.

We now build (propose) a ferrielectric state of PbZrO$_{3}$ preserving the largest distortion in the AFE$_{40}$ phase (that is, the $R_{4}^{+}$ mode) and imposing the Pb displacements to follow an \textit{up-up-down} pattern instead of the usual \textit{up-up-down-down} modulation (see Figures ~\ref{fig_crystal}{\bf c}, \ref{fig_crystal}{\bf d} and \ref{fig_crystal_schem}{\bf b}). The resulting state (`FiE' in the following) can be described by a 30 atom primitive cell and belongs to space group $Ima2$. The lattice vectors of the primitive cell are given by $\bb{b}_{1}=\bb{a}_{1}+2\bb{a}_{2}+\bb{a}_{3}$, $\bb{b}_{2}=-\bb{a}_{1}+\bb{a}_{2}$ and $\bb{b}_{3}=2\bb{a}_{3}$, where $\bb{a}_{1}=a(1,0,0)$, $\bb{a}_{2}=a(0,1,0)$, $\bb{a}_{3}=a(0,0,1)$ are the lattice vectors of the cubic perovskite cell and $a$ the cubic lattice parameter.  The FiE phase presents uncompensated dipoles by construction, and is allegedly the simplest ferrielectric phase of PbZrO$_{3}$ compatible with the leading $R_{4}^{+}$ instability one can imagine.

By performing an \textit{ab-initio} structural optimization we find that the proposed FiE phase is a stable polymorph of PbZrO$_{3}$. In the relaxed structure the three symmetry-inequivalent Pb atoms are displaced by $-$0.21~\AA\, $+$0.16~\AA\ and $+$0.23~\AA\ along $[1\bar{1}0]$ with respect to the cubic parent structure, yielding an in-plane polarization of 0.11~C m$^{-2}$ along the same direction. The dominant distortions with respect to the cubic reference are (i) the $R_{4}^{+}$ mode, accounting for 53.2\% of the total (see Supplementary Figure 2{\bf a} and Supplementary Movie~1), (ii) a $\Sigma_{2}$ mode at $\bb{q}=(\frac{1}{3},\frac{1}{3},0)$ involving modulated Pb displacements along $[1\bar{1}0]$ and accounting for 33.8\% of the total distortion (out of every three Pb atoms, two displace by $+$0.16~\AA\ and one by $-$0.32~\AA\ according to this mode, so they create no net dipole; see Supplementary Figure 2{\bf b} and Supplementary Movie~2), and (iii) a $\Gamma_{4}^{-}$ polar distortion, which involves opposed Pb and O displacements along $[1\bar{1}0]$ and accounts for 8.8\% of the total distortion (see Supplementary Figure 2{\bf c} and Supplementary Movie~3). This last distortion is responsible for the onset of a net polarization. All the remaining distortions combined amount to 4.2\% of the total. The Wyckoff positions of the optimized structure are given in Supplementary Table~1. The resulting state shows notorious resemblance with the AFE$_{40}$ phase and can be considered a ferrielectric variant of it.

We now compare the relative energies of the FiE phase (using the PBEsol exchange-correlation potential, see Methods) with those of (i) the commonly accepted ground state (AFE$_{40}$), (ii) an antipolar phase obtained from a soft-mode condensation of AFE$_{40}$ (AFE$_{80}$)~\cite{baker2021re}, which presents space group $Pnma$, and (iii) a low energy ferroelectric phase with $R3c$ symmetry (`FE' in the following). (This FE phase is the ground state of BiFeO$_3$ and features a spontaneous polarization along the [111] direction and an $a^-a^-a^-$ oxygen octahedra tilt pattern in Glazer's notation~\cite{dieguez11}.) We find that the four studied polymorphs lie within an energy range of approximately 1~meV per f.u., the phase hierarchy being, from most to least stable, AFE$_{80}$, FiE , AFE$_{40}$, and FE. The energies relative to the AFE$_{40}$ state are $-$0.89~meV per f.u, $-$0.84~meV per f.u., 0~meV per f.u, and $+$0.23~meV per f.u., respectively. Given the small energy differences, we next consider the effect of the exchange-correlation potential in the polymorph hierarchy.

{\bf Effect of exchange-correlation potential and volume in the DFT energies.} In order to verify the robustness of our results against the choice of the exchange-correlation functional, we repeat the calculations using the local density approximation (LDA), the Perdew, Burke and Ernzerhof implementation of the generalized gradient approximation (PBE)~\cite{PBE}, and the recently proposed strongly constrained and appropriately normed (SCAN) meta-GGA functional~\cite{sun15}. The Kohn-Sham energies are displayed in Table~\ref{tab_pzoenergies}. The scenario described by LDA is similar to that of PBEsol, the most notable difference being that the FiE state is now more stable than the AFE$_{80}$ phase. By contrast, the calculations with PBE predict the most stable phase to be the FE polymorph by more than 5~meV per f.u., followed by the AFE$_{40}$ and FiE states. Finally, SCAN predicts the AFE$_{40}$ phase to be the ground state, very closely followed by the AFE$_{80}$ phase, and then by the FiE and FE phases. Interestingly, we find that SCAN predicts both antiferroelectric phases to be independent minima of the energy, albeit with a very small barrier between them (see Supplementary Figure 3).

\begin{figure}
	\includegraphics[scale=0.37]{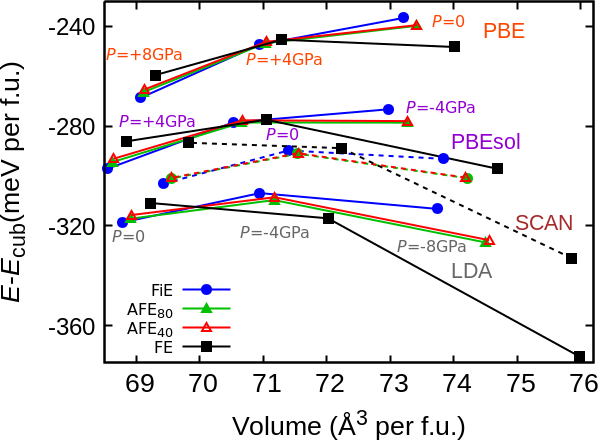}
	\caption{{\bf Kohn-Sham energies as a function of volume per formula unit for the PbZrO$_3$ polymorphs at different pressures and with different exchange-correlation functionals.} LDA results for 0, $-$4 and $-$8~GPa are displayed (bottom). PBEsol and SCAN results shown for $+$4, 0 and $-$4~GPa (centre, solid and dashed lines, respectively). PBE results for $+$8,	$+$4 and 0~GPa shown (top). The energies are shown relative to the energy of the cubic phase for each pressure. Blue circles, red empty triangles, green filled triangles, and black squares correspond to the FiE, AFE$_{40}$,	AFE$_{80}$, and FE phases, respectively. The results for SCAN are shifted by $+$20~meV per f.u. for clarity. In the used convention, a positive external pressure implies a compression.
	}
	\label{fig_volume}
\end{figure}

 The choice of exchange-correlation functional is known to affect the optimized volume in a DFT calculation. LDA (PBE) typically overbinds (underbinds) the system and thus tends to give relatively small (large) cell volumes as compared to experiments. The volumes obtained with PBEsol and SCAN~\cite{sun16} are in general between those of LDA and PBE, and typically closer to experiment. Hence, the variations in the polymorph hierarchy with the exchange-correlation functional could be mostly due to a volume effect. In order to test this hypothesis, we perform structural optimizations of the PbZrO$_3$ polymorphs under constant pressure for the four exchange-correlation functionals. First, we find that the equilibrium lattice parameter of the cubic perovskite phase given by LDA at a tensile pressure of $-$4~GPa ($-$8~GPa) is very close to that of PBEsol and SCAN (PBE) at zero pressure. We thus optimize the four PbZrO$_3$ polymorphs at $-$4~GPa and $-$8~GPa with LDA, at $-$4~GPa and $+$4~GPa with PBEsol and SCAN, and at $+$4~GPa and $+$8~GPa with PBE. The results are displayed in Figure~\ref{fig_volume}. At smaller cell volumes the most stable phase is FiE, followed by AFE$_{80}$, AFE$_{40}$, and finally FE. On the other end, at large cell volumes the FiE phase becomes the least stable and FE is the most favourable polymorph. It is thus clear that the cell volume plays an important role in the relative stability of these states. Also, it is apparent that the variations in the relative polymorph energies among the different exchange-correlation functionals can be largely attributed to a volumetric effect.

 \begin{table}
 \centering
 \begin{tabular}{lcccc}
PbZrO$_3$   &   LDA             & PBEsol            & PBE        & SCAN \\ \hline 
 FiE        & $-$2.96           & $-$0.84           & $+$2.87    & $+$1.08 \\  
 AFE$_{80}$ & $-$1.46           & $-$0.89           & $-$0.29    & $+$0.21 \\  
 AFE$_{40}$ & $\phantom{-}$0.00 &$\phantom{-}$0.00  & $\phantom{-}$0.00  & $\phantom{-}$0.00\\  
 FE         & $+$4.81           & $+$0.23           & $-$8.78  & $+$2.01 \\               
    \end{tabular}
 \begin{tabular}{lcccc}
 \\
PbHfO$_3$  &   LDA             & PBEsol            & PBE         & SCAN \\\hline 
FiE        & $-$0.78           & $+$0.31           &$+$2.61    & $+$3.53 \\    
AFE$_{80}$ & $-$0.77           & $-$0.50           &$-$0.10    & $+$0.18 \\    
AFE$_{40}$ & $\phantom{-}$0.00 & $\phantom{-}$0.00 &$\phantom{-}$0.00  &$\phantom{-}$0.00\\
FE         & $+$2.76           & $-$0.23           &$-$6.75    & $-$5.44 \\    
    \end{tabular}
    
	 \caption{Kohn-Sham energies in meV per formula unit (f.u.) of the studied phases in PbZrO$_3$ (above) and PbHfO$_{3}$ (below) relative to the AFE$_{40}$ phase.
	 The different columns correspond to different choices of the exchange-correlation potential.}
	 \label{tab_pzoenergies}
\end{table}

{\bf Comparison with lead hafnate.}  Many experimental works have reported the low temperature antiferroelectric ground state of lead hafnate to be isostructural to the AFE$_{40}$ phase of PbZrO$_3$, using X-ray diffraction~\cite{forker1973,corker1998investigation,PHO_thinfilms}, neutron diffraction~\cite{corker1998investigation,madigout1999crystallographic,fujishita2008analysis}, electron microscopy~\cite{madigout1999crystallographic,bussmann2015phase,fan2020tem} and Raman spectroscopy~\cite{PHO-Raman,jankowska1995raman}. Hence, to further test the soundness of our conclusions for PbZrO$_{3}$, we compute the Kohn-Sham energies of the four phases here considered for PbHfO$_{3}$, using the LDA, PBEsol, PBE and SCAN functionals. The results are displayed in Table~\ref{tab_pzoenergies}. We find that for LDA the polymorph hierarchy is exactly the same as that of PbZrO$_3$, the FiE phase being the most stable, closely followed by AFE$_{80}$, AFE$_{40}$ and finally FE. The four phases lie within 3.5~meV per f.u. In contrast, for PBEsol the most stable phase is AFE$_{80}$, closely followed by the FE, AFE$_{40}$ and FiE states. In this case, the obtained energy difference between the most stable and the least stable is of only 0.8~meV per f.u. For PBE, as in the case of PbZrO$_3$, we find that the FiE is the highest energy polymorph and that the FE phase is predicted to be the most stable.
Interestingly, our SCAN calculations predict (similarly to PBE) the FE phase to be the most favourable phase of PbHfO$_{3}$, in clear contrast with the available experimental data (and with our SCAN predictions for PbZrO$_{3}$). Besides this, the polymorph hierarchy dependence of PbHfO$_{3}$ on the exchange-correlation functional qualitatively follows the same trends as in PbZrO$_{3}$. For example, our simulations predict the AFE$_{80}$ state to be marginally more favourable than the AFE$_{40}$ phase  (except with SCAN).
The calculations for PbHfO$_{3}$ are also in line with the results for PbZrO$_3$ as regards the obtained unit cell volumes: LDA yields the smallest ones, which favours the FiE phase over the rest. We also note that our simulations predict the AFE$_{80}$ phase of PbHfO$_3$ to be more stable than the AFE$_{40}$ phase by less than 1~meV per f.u. (except with SCAN, for which the AFE$_{40}$ is more stable, as in the case of PbZrO$_{3}$), in agreement with recent calculations~\cite{baker2021re}. Hence, overall, the trends found for PbZrO$_3$ for the polymorph hierarchy dependence on the exchange-correlation (and, in turn, the cell volume) are also found in PbHfO$_3$.

{\bf Phonon bands and zero-point energies.} Noting that the energy differences between competing polymorphs are tiny, we turn our attention to the zero-point contributions (rarely considered in DFT investigations of ferroelectrics) to see whether they may have an impact.
 
\begin{figure}
 \includegraphics[scale=0.13]{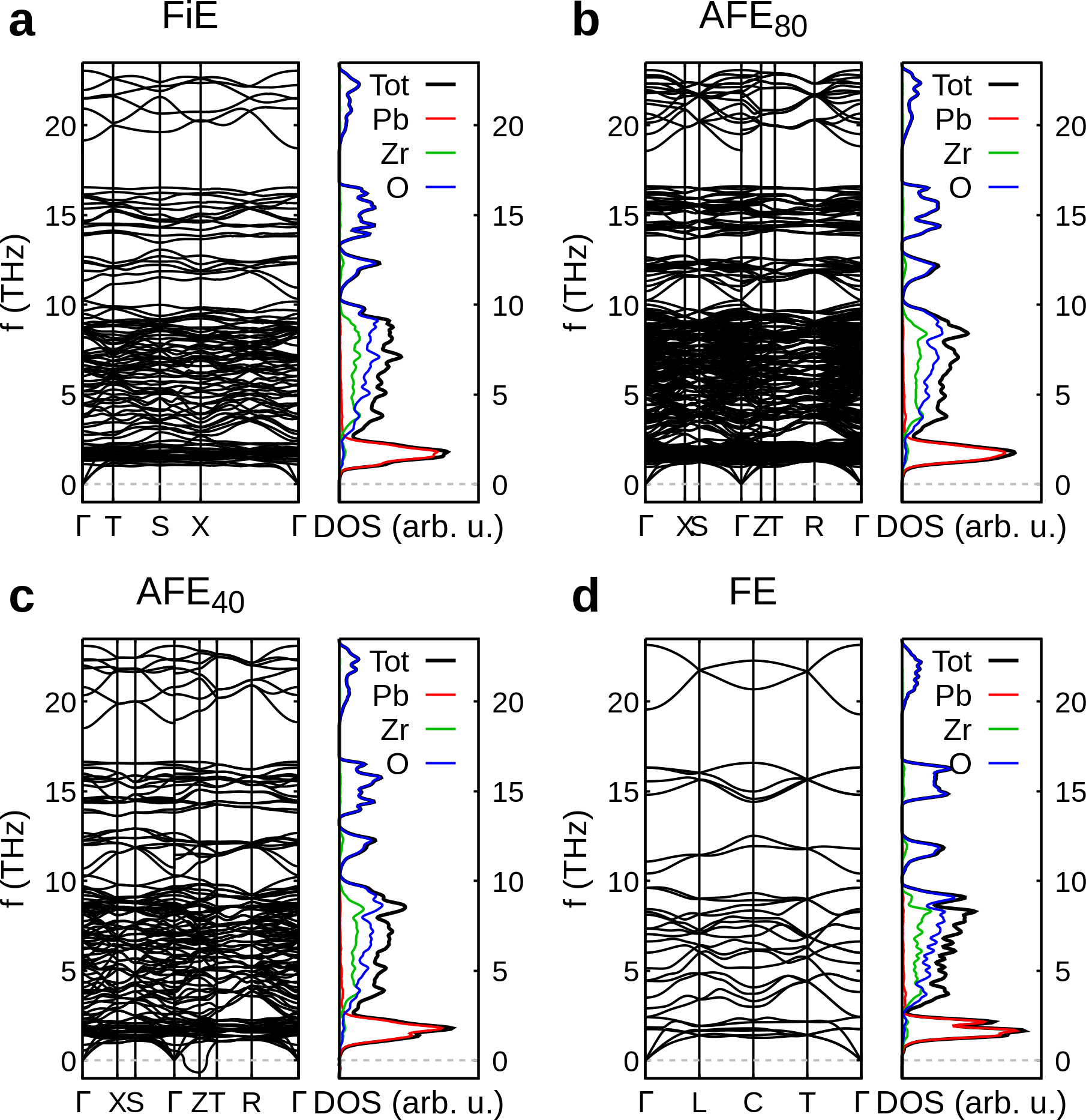}
	\caption{{\bf Phonon band structure and phonon density of states of the studied PbZrO$_{3}$ phases computed with PBEsol.} The results for the FiE , AFE$_{80}$, AFE$_{40}$, and FE states are shown in panels {\bf a}, {\bf b}, {\bf c}, and {\bf d}, respectively.}
	\label{fig_phonons}
\end{figure}

We compute the phonon dispersion and phonon density of states (DOS) for the four phases using first principles (see Methods). The results using PBEsol are displayed in Figure~\ref{fig_phonons}. We see that the FiE phase (as well as AFE$_{80}$ and FE) shows no instabilities. Only the AFE$_{40}$ state shows an unstable phonon branch in the vicinity of the $Z$ point. It is precisely by condensing this instability that we found the AFE$_{80}$ polymorph, in the same way as done in Ref.~\onlinecite{baker2021re}. Besides this, the phonons of the four phases show similar features: relatively flat bands associated to Pb around 2~THz, a band dominated by Zr and O between 2~THz and 10~THz, and high frequency bands associated to O up to almost 24~THz. It is worth noting that we find no instabilities in the AFE$_{40}$ phase with SCAN (see Supplementary Figure 4). This is consistent with both phases being two separate energy minima, as discussed before. We will thus not pursue the AFE$_{80}$ phase with SCAN in the following, since SCAN predicts it to be less stable than the AFE$_{40}$ polymorph.
 
The phonon calculations allow us to compute the zero-point energies ($E_\mathrm{ZPE}$) (see Methods); our PBEsol results are listed in the third column of Table~\ref{tab_dftenergies}. The difference among them is smaller than 2~meV per f.u. Yet, the AFE$_{80}$ phase shows a larger $E_\mathrm{ZPE}$ than the FiE state, the difference being big enough to invert the energy hierarchy (see column four in Table~\ref{tab_dftenergies}). Thus, our calculations predict that the zero-point energy stabilizes the FiE over the AFE$_{80}$; hence, according to our PBEsol results, PbZrO$_3$ would be a `quantum ferrielectric'.
 
We note that the few unstable modes of AFE$_{40}$ in the vicinity of the $Z$ point are excluded from the zero-point energy summations with LDA, PBEsol and PBE; thus, the zero-point energy of the AFE$_{40}$ phase is slightly underestimated with these functionals and its stability is overestimated (see Methods).
 
We also study the effect of the exchange-correlation functional on the zero-point energies. The results for LDA, PBE and SCAN are listed in Table~\ref{tab_dftenergies} in columns five to seven, eight to ten and eleven to thirteen, respectively. In LDA, PBE and SCAN the zero-point energies do not alter the phase hierarchy given by the Kohn-Sham energies. 
Consequently, considering the zero-point correction, LDA predicts PbZrO$_3$ to be ferrielectric (but not a quantum ferrielectric), PBE predicts the FE phase to be the most stable and SCAN predicts the AFE$_{40}$ to be the ground state.

Overall, our simulations draw three possible scenarios for PbZrO$_3$ at zero Kelvin: LDA and PBEsol predict the FiE state to be the most stable phase at low temperatures, PBE predicts FE to be the ground state, while the AFE$_{40}$ phase is the most favourable one according to SCAN. Surprisingly, our calculations reveal that the AFE$_{80}$ phase is not the most stable solution with any of the four functionals. In order to elucidate the most feasible scenario, a study of the temperature dependence of relative phase stability is warranted. 

\begin{table*}
 \centering
 \begin{tabular}{lcccccccccccccccc}
     & &\multicolumn{3}{c}{PBEsol}& &\multicolumn{3}{c}{LDA}& &\multicolumn{3}{c}{PBE}& &\multicolumn{3}{c}{SCAN}\\
  & & $E_\mathrm{KS}$    & $E_\mathrm{ZPE}$ & $E_\mathrm{tot}$& &$E_\mathrm{KS}$    &$E_\mathrm{ZPE}$ & $E_\mathrm{tot}$& & $E_\mathrm{KS}$    &$E_\mathrm{ZPE}$ & $E_\mathrm{tot}$& & $E_\mathrm{KS}$    &$E_\mathrm{ZPE}$ & $E_\mathrm{tot}$ \\ \hline 
 FiE       & &   $-$0.84         & 250.68    & 249.84 & &  $-$2.96          & 257.63    & 254.67 & &  $+$2.87          & 242.17   & 245.04   & & $+$1.08           & 255.76 & 256.84 \\           
 AFE$_{80}$& &   $-$0.89         & 250.92    & 250.03 & &  $-$1.46          & 257.73    & 256.27 & &  $-$0.29          & 242.70   & 242.41   & & $+$0.21           &  -     & - \\           
 AFE$_{40}$& & $\phantom{-}$0.00 & 250.87    & 250.87 & &$\phantom{-}$0.00  & 257.62    & 257.62 & & $\phantom{-}$0.00 & 242.68   & 242.68   & & $\phantom{-}$0.00 & 256.36 & 256.36   \\           
 FE        & &   $+$0.23         & 251.28    & 251.51 & &  $+$4.81          & 257.58    & 262.39 & &  $-$8.78          & 244.17   & 235.39   & & $+$2.01           & 256.13 & 258.14   \\    
    \end{tabular}
	 \caption{Computed energies (in meV per formula unit) of the studied PbZrO$_3$ phases using PBEsol (columns two to four), LDA (columns five to seven), PBE (columns eight to ten) and SCAN (columns eleven to thirteen). For each exchange-correlation functional we display the Kohn-Sham energies relative to 
	 the AFE$_{40}$ phase (first column of each block), zero point energy (second column of each block), and sum of Kohn-Sham energies
	 and zero point energies ($E_\mathrm{tot}=E_\mathrm{KS}+E_\mathrm{ZPE}$, third column of each block). The zero point energy for the AFE$_{80}$ phase was not considered with SCAN (see text).}
	 \label{tab_dftenergies}
\end{table*}

\begin{figure*}
	\includegraphics[scale=0.22]{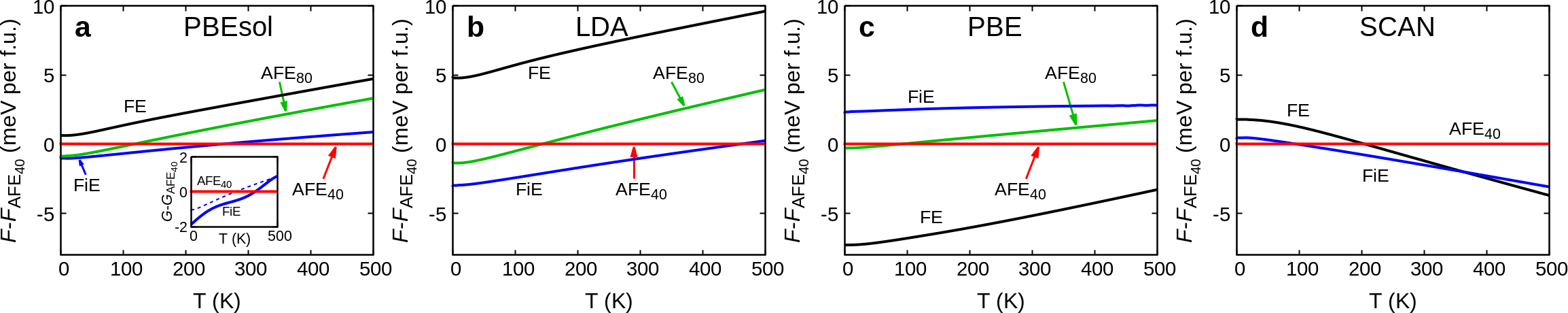}
	\caption{{\bf Free energies of the studied PbZrO$_{3}$ phases as a function of temperature.} Free energy differences to the AFE$_{40}$ phase (red) as a function of temperature for the FiE (blue), AFE$_{80}$ (green), and FE (black) phases of PbZrO$_3$ are shown in  panels {\bf a}, {\bf b}, {\bf c} and {\bf d} for PBEsol, LDA, PBE, and SCAN, respectively. The inset in {\bf a} shows the result for the FiE and AFE$_{40}$ states when accounting for thermal expansion (i.e., zero pressure conditions) as obtained with PBEsol. The dashed line in the inset corresponds to the free energy of the FiE without considering the thermal expansion.}
	\label{fig_freeenergy}
\end{figure*}

{\bf Temperature dependence of the free energies.} Having completed the DFT investigation of these polymorphs at 0~K, we now try to get information about how their relative stability evolves with temperature. For that, we estimate the temperature dependence of the corresponding free energies within the harmonic approximation (see Methods). In Fig.~\ref{fig_freeenergy}{\bf a} we display the free energy difference between the PbZrO$_3$ polymorphs and the AFE$_{40}$ phase, as computed with the PBEsol functional. At low temperatures the FiE phase is the most stable, and remains so up to 255~K, where a transition to the AFE$_{40}$ phase occurs. We also find that at 115~K the vibrational entropy stabilizes the AFE$_{40}$ phase over AFE$_{80}$. The FE state remains a high-energy polymorph in the studied temperature range.
 
Let us recall that the unstable phonons near the $Z$-point in the AFE$_{40}$ phase are not included in the free energy summation. While in the zero-point energy calculation this exclusion results in an overestimation of the stability of the AFE$_{40}$ state, in this case we are slightly underestimating it. Therefore, these errors can be expected to partially cancel each other. (Note that with SCAN this phase shows no unstable phonons and therefore this comment does not apply to that case). At any rate, the number of phonons excluded in the summations is very small and its impact on the free energies is expected to be minor. Also, we do not consider the cubic perovskite phase in this comparison since it presents many strong instabilities spanning throughout the whole Brillouin zone (see Supplementary Figure 1), and therefore we do not expect the harmonic approximation to the free energy to yield meaningful results for that polymorph.

We further compute the effect of the exchange-correlation potential in the free energy evolution of the four polymorphs. Figures~\ref{fig_freeenergy}{\bf b},  \ref{fig_freeenergy}{\bf c} and \ref{fig_freeenergy}{\bf d}, show the results for LDA, PBE and SCAN, respectively. The transition between the AFE$_{40}$ and AFE$_{80}$ phases occurs in the temperature range between 80~K and 130~K for the three exchange-correlation functionals in which the AFE$_{80}$ has been considered. The scenario described by LDA is qualitatively identical to that of PBEsol, the crossing between the FiE and AFE$_{40}$ phases occurring at a higher temperature (460~K). PBE again paints a qualitatively different picture: the most stable phase among those studied turns out to be the FE one throughout the considered temperature range. The prediction of a FE phase as the ground state of PbZrO$_3$ up to 500~K is at odds with the current knowledge about this compound and cannot be reconciled with the experimental results (in particular, with the observed cubic to AFE$_{40}$ transition in the neighbourhood of 500~K)~\cite{shirane51,whatmore1979}.With SCAN the usual AFE$_{40}$ phase is predicted to be the most stable at low temperatures, but we find that the polar phases FiE and FE become the most stable ones at 100~K and 360~K, respectively. A polar phase more stable than the antiferroelectric state in a wide temperature range spanning from 500~K down to almost 100~K is also all but impossible to reconcile with the experiments.
Therefore, our findings suggest that both PBE and SCAN are not appropriate for studying the structural behaviour of PbZrO$_3$.
 
{\bf Effect of thermal expansion.} The (free) energy differences among the studied phases continue to be rather small. Further, we have seen above that the relative stability of the polymorphs of interest here depends critically on volumetric factors. Hence, we can expect the thermal expansion to have an impact on the relative stability between states. We can address this issue with first-principles calculations within the so-called quasi-harmonic approximation (see Methods), though at a high computational cost. In order to reduce the computational burden, we compute the zero-pressure Gibbs free energy for the FiE and AFE$_{40}$ phases only, which we display in the inset of Fig.~\ref{fig_freeenergy}{\bf a}. (The effect of the thermal expansion on the FE and AFE$_{80}$ phases is expected to follow a similar trend.) The most relevant outcome of considering the thermal expansion is a shift of the crossing temperature between the FiE and AFE$_{40}$ states to a higher value (370~K). This change is rather large, and can be understood considering that the energy differences between these phases are small, and that the dependence of the free energy and the Gibbs free energy with temperature is very similar for both polymorphs (recall that we are showing energy differences in the figure). At any rate, considering the thermal expansion yields the same qualitative scenario: the FiE phase remains the most stable at low temperatures, while the AFE$_{40}$ phase is the most favourable at high temperatures.
 
Incidentally, let us note that thermal expansion cannot be expected to reconcile the PBE results with experiments. As discussed above, PBE underbinds PbZrO$_{3}$, yielding larger volumes than PBEsol and LDA. At larger cell volumes the FE state becomes the most stable phase, and therefore accounting for the thermal expansion in the PBE calculations would most likely predict an even greater relative stability of the FE state. Regarding SCAN, assuming that the thermal expansion is similar to that observed with PBEsol, it would tend to favour the FiE phase over the AFE$_{40}$ polymorph. In this scenario the SCAN results would be even more incompatible with the experiments.

\pagebreak 

\vspace{5mm}{\bf Discussion}

{\bf Status of our results.}
Our prediction of a ferrielectric ground state for PbZrO$_3$ relies on energy differences that are small. One could
think that these differences are smaller than the expected accuracy of first-principles calculations. Yet, as explained in Methods, we have carefully checked that our calculations are well converged (beyond the energy differences between polymorphs) against the computational parameters controlling their precision ($\bb{k}$-point grid and plane-wave cut-off for Kohn-Sham energies, $\bb{q}$-point grid for zero-point energy and free energy calculations). We have also shown that our results are robust against the effect of the thermal expansion for the FiE and AFE$_{40}$ phases. Furthermore, we have explicitly tested our predictions against the choice of exchange-correlation functional, finding that the qualitative picture is robust for the LDA and PBEsol approximations, while the PBE and SCAN functionals fail to describe the low energy landscape of PbZrO$_3$. Finally, our calculations for the isostructural compound PbHfO$_3$ indicate a very similar behaviour and support the validity of our conclusions for PbZrO$_3$. Hence, we have reasons to believe that our results are qualitatively meaningful.

On the other hand, we find that small deviations in the computed free energy differences may result in significant shifts in the estimated transition temperatures. (See, e.g., the shift from 255~K to 460~K, for the FiE to AFE$_{40}$ transition, when moving from PBEsol to LDA.) Hence, we do not expect our quantitative results to be accurate.

Most importantly, let us stress that in this work we have considered only one ferrielectric polymorph, arguably the simplest one that can host the leading $R$-point instability of the cubic phase, finding that it displays a lower energy than the known antiferroelectric phases of PbZrO$_3$ and PbHfO$_3$. Yet, this is the only ferrielectric phase we have attempted; it is possible that other phases with uncompensated dipoles may display an even lower (free) energy. Hence, the phase diagram of PbZrO$_{3}$ might be even richer than the one discussed here.

{\bf Connection with experiment.}
Several techniques have been used by many experimental groups to obtain samples of PbZrO$_3$. These include the growth of crystals by the flux method~\cite{Sawaguchi1951,scott1972crystal,glazer1993structure,tagantsev2013origin}, film growth by the sol-gel method~\cite{garnweitner2005nonaqueous,wu2018defect,coulibaly2020enhancement}, chemical vapour deposition~\cite{moret2002structure}, atomic layer deposition~\cite{harjuoja2007atomic} and pulsed layer deposition~\cite{ikeda2000dielectric,chaudhuri2011epitaxial}. Even if this is not an exhaustive list, it is worth noting that all the cited works report preparation temperatures of at least 900~K (even with the atomic layer deposition method described in Ref.~\onlinecite{harjuoja2007atomic}, the authors have to apply a thermal treatment at 900~K to be able to obtain the perovskite phase). At the same time, the experimental transition temperature between the cubic and antiferroelectric phases is reported to be about 500~K~\cite{Sawaguchi1951,whatmore1979,fujishita2000temperature}. Hence, when preparing PbZrO$_3$ the system always starts from the cubic phase, and the AFE$_{40}$ phase is condensed upon cooling at around 500~K.
 
Our calculations estimate the FiE to AFE$_{40}$ transition temperature to be somewhere between 255~K and 460~K, in any case below the cubic to AFE$_{40}$ experimental transition. Hence, when the actual PbZrO$_{3}$ samples cool down from the growth temperature towards our predicted AFE$_{40}$ to FiE transition point, it is conceivable for this transformation to be kinetically hindered, provided that the available thermal energy is not sufficient to overcome the energy barrier separating the two free energy minima. Note that the FiE and AFE$_{40}$ (or AFE$_{80}$) states feature distinct order parameters and cannot be connected by a continuous symmetry breaking; hence, the transition between them must be a first-order discontinuous one. (Unlike the case of the AFE$_{80}$ phase, which stems from the condensation of a single soft mode of the AFE$_{40}$ parent structure.) More specifically, in such a discontinuous phase transition, the anti-polar \textit{up-up-down-down} pattern of Pb displacements must get undone and replaced by the \textit{up-up-down} uncompensated arrangement of the FiE state. According to Ref.~\onlinecite{iniguez2014first}, the condensation of the Pb displacements in the \textit{up-up-down-down} pattern lowers by around 100~meV per f.u. the energy of the structure with $R$-point octahedral tilts condensed. This energy reduction is, in essence, the barrier the system needs to overcome in order to escape the AFE$_{40}$ minimum and transform into FiE. (The actual lowest-energy transition path will present a lower barrier, but the mentioned value should give us a meaningful order of magnitude.) For comparison, in PbTiO$_3$ the condensation of the polar ferroelectric distortion (characterized by a similar off-centring of Pb atoms) lowers the energy by 83~meV per f.u. according to our PBEsol calculations; and this estimated energy barrier for escaping the ferroelectric state (also an upper bound) corresponds to an experimental transition temperature of 768~K~\cite{bhide62}. Hence, a comparable energy barrier in a similar compound is overcome only at a very high temperature, much higher than the predicted FiE to AFE$_{40}$ transition point. Thus, this suggests that the transition from AFE$_{40}$ to FiE may indeed be kinetically hindered. In this scenario, our calculations are compatible with the AFE$_{40}$ phase being the widely observed experimental structure of PbZrO$_3$ at all temperatures below 500~K.

Has the FiE state of PbZrO$_{3}$ ever been stabilized by application of an electric field? Several works have reported double hysteresis loops for this compound at room temperature~\cite{zhai03,ge14,zhang16,nguyen20,milesi21}, which is one of the experimental signatures of an antiferroelectric material. An electric field triggers a transition to a polar state, and in most of the reported loops PbZrO$_3$ returns to an antipolar state (with zero remnant polarization) upon releasing the field. This seems to suggest that subjecting the samples to a field does not stabilize the FiE phase. One possible reason for this could be that the actual transition point between the AFE$_{40}$ and FiE phases is below room temperature; if this is the case, the recovery of the AFE$_{40}$ state upon removal of the field, instead of FiE, is to be expected. Note that this possibility is compatible with our first-principles results, given the large error bar associated to the estimated FiE--AFE$_{40}$ transition temperature (see discussion above). 
Alternatively, it is also possible that, even if the FiE phase were to be most stable state at room temperature, the energy barrier between the polar and antipolar states might be lower than that separating the polar and FiE phases, resulting in a preferred back-switch to the antipolar AFE$_{40}$ state.

Having said this, we should note that the authors of Ref.~\onlinecite{burkovskyferrielectric} report the stabilization under an applied field of a phase that might be ferrielectric (where the Pb displacements show modulation of a longer period than that of our FiE phase). Additionally, Refs.~\onlinecite{pintilie2008coexistence} and \onlinecite{guo19} report pinched hysteresis loops for thin films of PbZrO$_3$ at room temperature, which are compatible with the system reaching a FiE state at zero field. The remnant polarizations observed by these authors are right below 0.10~C m$^{-2}$, which is very close to the ferrielectric polarization we compute (0.11 C m$^{-2}$). Thus, we cannot rule out the possibility that the FiE phase might indeed be stable at ambient conditions, and that maybe it has already been obtained experimentally but not recognised as such.

As regards the AFE$_{80}$ phase, our first-principles calculations predict that it should appear as a phonon instability of the AFE$_{40}$ state somewhere between 115~K and 140~K (except with SCAN, which predicts that the AFE$_{80}$ phase is a separate energy minimum and less favourable than the AFE$_{40}$ state). Yet, to our knowledge, there is no experimental evidence supporting the occurrence of the AFE$_{80}$ phase. The distortions involving the condensation of this state are not very small (according to our calculations the largest atomic displacement between the phases is around 0.15~\AA); yet, the AFE$_{40}$ and AFE$_{80}$ structures are overall very similar, so it is possible that this phase may have been obtained experimentally but not identified as such.

Another possibility is that the occurrence of the AFE$_{80}$ phase is precluded by lattice quantum fluctuations. These effects stem from the wave-like nature of the atoms, go beyond the zero-point corrections here considered, and are known to have a significant impact in the phase diagram of related perovskites like BaTiO$_{3}$~\cite{iniguez02} and SrTiO$_{3}$~\cite{zhong95b}. Quantum fluctuations favour the low-temperature stability of high-temperature states and are, for example, responsible for the `quantum paraelectric' nature of SrTiO$_{3}$~\cite{zhong95b,muller79}. They can be expected to reduce the temperature of the AFE$_{40}$ to AFE$_{80}$ transition, and might even suppress it completely.

{\bf Intermediate polar phase.}
The existence of an intermediate phase between the cubic and AFE$_{40}$ states of PbZrO$_{3}$ was already observed in the first experimental works on this compound~\cite{shirane51}, yet its structure remains controversial~\cite{liu18}. This phase could be ferroelectric~\cite{tennery1965study,scott1972crystal} (although some authors claim it could be antiferroelectric~\cite{fujishita84,teslic1998atomic}), possibly with rhombohedral symmetry~\cite{tennery66,whatmore1979}. Some studies have found this structure to be stable between 479~K and 505~K~\cite{scott1972crystal}, while in others the observed stability region is between 503~K and 506~K~\cite{whatmore1979} - at any rate it is clear that the stability window is narrow.

Do our results offer any information about this intermediate phase? On the one hand, the results obtained with LDA and PBEsol (which we believe are the most reliable) seem to rule out the rhombohedral FE state here considered as a possible candidate, as the computed free energies are much higher than those of competing polymorphs (see Figs.~\ref{fig_freeenergy}{\bf a} and ~\ref{fig_freeenergy}{\bf b}). On the other hand, our PBE and SCAN results predict that the FE (and even the FiE) state may indeed be stable at high temperatures (see Figs.~\ref{fig_freeenergy}{\bf c} and ~\ref{fig_freeenergy}{\bf d}), which would, in principle, draw a more optimistic scenario than our LDA and PBEsol results. However, as already mentioned, the PBE and SCAN results seem all but incompatible with the experimental evidence on PbZrO$_{3}$ (i.e., they penalize the well-known AFE$_{40}$ state at room temperature); hence, we do not think they provide us with reliable data to discuss the nature of the intermediate state.

{\bf Stabilization of the FiE phase with pressure.}
Our calculations with different exchange-correlation functionals and with pressure indicate that compression can further stabilize the FiE phase over the AFE$_{40}$ polymorph. Hence, it is possible that applying pressure on PbZrO$_3$ samples at room temperature may help induce a AFE$_{40}$ to FiE transition (provided the kinetic barrier can be overcome), and that the ferrielectric state may remain stable upon releasing the pressure. Note that some works have already reported the stabilization of polar phases in PbZrO$_3$ over the antipolar AFE$_{40}$ under pressure~\cite{prosandeev2014anomalous}, or a polar modulated phase with higher density than the antiferroelectric phase~\cite{wei2020unconventional}, which could be related to the findings presented here.
  
{\bf Implications for effective theories.}
Finding an explanation for the occurrence of the AFE$_{40}$ phase of PbZrO$_{3}$, in terms of effective atomistic theories or Ginzburg-Landau field models, is a challenge that remains not fully resolved. A key step forward was the demonstration, in 2014, of the all-important role played by the antiphase rotations of the oxygen octahedra, by means of both vibrational spectroscopy~\cite{hlinka14} and DFT calculations~\cite{iniguez2014first}. That the octahedral tilts are the dominant instability of PbZrO$_{3}$'s cubic phase is in fact apparent; it can be immediately appreciated by inspecting the structure of the antipolar state. However, the tilts had been traditionally regarded non-essential~\cite{tagantsev2013origin} and the focus had been, almost exclusively, on the antipolar Pb displacements. In contrast, Refs.~\onlinecite{hlinka14,iniguez2014first} showed that the tilts (i.e., the $R_{4}^{+}$ mode discussed above) are key to stabilize the antipolar pattern ($\Sigma_{2}$ mode) over competing polar orders, thanks to a trilinear coupling that also involves a third (auxiliary) distortion (a mode with $S_{4}$ symmetry at $\bb{q}=(\frac{1}{4},\frac{1}{4},\frac{1}{2})$)~\cite{iniguez2014first}.

Another step forward was the elucidation~\cite{patel2016atomistic} of the atomistic couplings responsible for the mixing of the phonon bands corresponding to the polar (zone-centre) and tilting (zone-boundary) instabilities, critical for the occurrence of the low-energy soft mode with $\Sigma_{2}$ symmetry. (Recall that this mode combines antipolar Pb displacements with incomplete octahedral tilts.) Based on these insights, an effective Hamiltonian has been recently proposed~\cite{xu19} and shown to reproduce the AFE$_{40}$ ground state and its relative stability against well-known competing polymorphs (e.g., the FE phase considered in this work). Hence, in principle, this model seems to capture the physics of tilts and (anti)polar distortions in PbZrO$_{3}$, at least in what regards the states commonly known.

However, our present results imply that the problem is even more difficult than we thought. Indeed, we have shown that the DFT ground state of PbZrO$_{3}$ combines antiphase tilts ($R_{4}^{+}$ mode), modulated polar distortions ($\Sigma_{2}$ band), and homogeneous polar distortions ($\Gamma_{4}^{-}$ mode). Note that this introduces a level of complexity not present in the AFE$_{40}$ state, which is characterized by only two main modes, $R_{4}^{+}$ and $\Sigma_{2}$. Determining whether the FiE ground state can be predicted by the effective Hamiltonian introduced in Ref.~\onlinecite{xu19}, or whether it will require the consideration of additional couplings, remains for future work.

To give a better feeling of the complexity inherent to the FiE state, let us note that this polymorph features modes associated to the $\bb{q}=(\frac{1}{6},\frac{1}{6},\frac{1}{2})$ wave vector ($S_{3}$ and $S_{4}$ symmetries), which occur as a by-product of combining the ferrielectric ($\Sigma_{2}$) and tilting ($R_{4}^{+}$) distortions. These secondary $S$ modes constitute a small contribution to the distortion connecting the cubic and FiE phases, amounting to less than 4\% of the total. Nevertheless, if we artificially remove them from the ground state structure, we find that the DFT energy increases by 25~meV per f.u. (In the case of the AFE$_{40}$ phase, a comparable increase of 27~meV per f.u. is obtained if the secondary $S_{4}$ mode is artificially removed~\cite{iniguez2014first}.) Importantly, this energy gain would be sufficient to completely alter the relative stabilities of the PbZrO$_{3}$ polymorphs considered here, as it would make the FiE state the least stable one at any temperature. Thus, if we want to construct a theory that captures the FiE ground state correctly, it will be critical to consider couplings between modes at $\bb{q}=(0,0,0)$, $\bb{q}=(\frac{1}{3},\frac{1}{3},0)$, $\bb{q}=(\frac{1}{2},\frac{1}{2},\frac{1}{2})$ and $\bb{q}=(\frac{1}{6},\frac{1}{6},\frac{1}{2})$.

Along these lines: using standard symmetry-analysis tools~\cite{hatch03,isodistort}, one can find that the $R_{4}^{+}$, $\Sigma_{2}$ and $S_{4}$ modes present in the FiE state are coupled via a trilinear interaction analogous to the $Q(R_{4}^{+})Q(\Sigma_{2})Q(S_{4})$ invariant discussed in Ref.~\onlinecite{iniguez2014first} in connection to the AFE$_{40}$ state. (Here $Q(R_{4}^{+})$ stands for the amplitude of the $R_{4}^{+}$ order parameter as it would appear in a Landau potential; {\em idem} for $Q(\Sigma_{2})$ and $Q(S_{4})$.) Thus, interestingly, the FiE and AFE$_{40}$ polymorphs are similar from this perspective. However, we find that the FiE state is further affected by a coupling of the form $Q(\Gamma_{4}^{-})(Q(\Sigma_{2}))^{3}$, which exists for $\Sigma_{2}$ modes at $\bb{q}=(\frac{1}{3},\frac{1}{3},0)$ and is probably key to the stability of FiE over AFE$_{40}$ (where no such interaction is active). While a detailed investigation of these aspects remains for future work, these observations do suggest that specific 4$^{\rm{th}}$-order interactions, between the polar and tilt degrees of freedom, should be part of an effective theory of PbZrO$_{3}$ that captures the FiE ground state. Thus, for example, a revision of the model in Ref.~\onlinecite{xu19} will likely be needed.

Last but not least, let us note that a satisfactory description of PbZrO$_{3}$'s AFE$_{40}$ state in terms of a field (Ginzburg-Landau) theory remains to be accomplished. This poses specific challenges, such as the need to treat, simultaneously, the polarization and tilt fields and their interactions. In particular, we can expect that complex high-order couplings, involving the gradients of both fields, will be required to obtain the $\Sigma_{2}$ instability dominating over other (anti)polar variants. Recent works have set the methodological basis to develop such a theory, and to identify (and compute) the relevant couplings from first principles~\cite{schiaffino17}. Obviously, the present discovery of a FiE ground state makes this task even more daunting; on the bright side, it provides us with a great case study where a first-principles-based field-theoretic method can make a definite difference over the traditional phenomenological approaches.

\vspace{5mm}{\bf Conclusion}
In this work we have discussed a low-energy ferrielectric phase of PbZrO$_3$. Our first-principles calculations indicate that our guessed structure is the most stable state among the known polymorphs of this compound; in fact, we predict this ferrielectric polymorph to be the actual ground state at low temperature, and to remain dominant (probably beyond room temperature) before transforming into the well-known antiferroelectric phase. We argue that our findings can be reconciled with the existing experimental picture (i.e., that PbZrO$_{3}$ is antiferroelectric -- not ferrielectric -- at ambient conditions and lower temperatures) provided that the transition to the ferrielectric state is kinetically hindered, which seems a reasonable scenario. In addition, our work indicates that the ferrielectric phase can be favoured over the antiferroelectric one by application of pressure. We hope the present results will give a new impulse to the investigation of the complex structural behaviour of PbZrO$_{3}$ (and analogous compound PbHfO$_{3}$) and the physical underpinnings of antiferroelectric and ferrielectric states.

\vspace{5mm}{\bf Methods}

{\bf Density functional theory calculations.}
We perform density functional theory (DFT) calculations with the Vienna \textit{Ab-initio} Simulation Package (\textsc{vasp})~\cite{VASP1,VASP2} to optimize the structure of the PbZrO$_3$ and PbHfO$_3$ phases. We use four different flavours of the exchange-correlation functional: the local density approximation (LDA), the Perdew, Burke and Ernzerhof implementation of the generalized gradient approximation (PBE), its revised version for solids (PBEsol)~\cite{PBEsol} and the recently proposed strongly constrained and appropriately normed (SCAN) meta-GGA functional~\cite{sun15}. We employ the plane augmented wave (PAW) method to represent the ionic cores, treating explicitly the following electronic orbitals: $5d$, $6s$, and  $6p$ for Pb, $4s$ $4p$ $4d$ and $5s$ for Zr, $5p$ $5d$ and $6s$ for Hf, and $2s$ and $2p$ for O. The electronic wave functions are represented in a plane-wave basis with a cut off of 500~eV. The electronic Brillouin zone (BZ) integrals are performed using Monkhorst-Pack~\cite{monkhorstpack} $\bb{k}$-point meshes of 5$\times$5$\times$5, 8$\times$4$\times$3, 8$\times$4$\times$6, and 8$\times$8$\times$8 for the FiE, AFE$_{80}$, AFE$_{40}$, and FE phases, respectively. The unit cells considered for the FiE, AFE$_{80}$, AFE$_{40}$, and FE phases have 30, 80, 40, and 10 atoms, respectively. We allowed the structures to relax until forces became smaller than 0.001~eV \AA$^{-1}$\ and stresses became smaller than 0.01~kBar. We checked that these calculation parameters yield well-converged
results. The polarization was obtained using the Berry phase approach within the modern theory of the polarization~\cite{modernTpol}.

{\bf Symmetry analysis.}
We use the \textsc{isodistort} tool~\cite{isodistort} within the crystallographic Isotropy suite~\cite{stokes06} to obtain the symmetry-adapted distortions connecting the daughter phases to the parent cubic perovskite phase. We also use the Invariants tool~\cite{hatch03} within the same suite to obtain the allowed invariant polynomials in the order parameter expansion of the Landau free energies.

{\bf Phonon, zero-point energy and free energy calculations.}
We compute the phonon dispersions and densities of states (DOSs) using the direct supercell approach implemented in the \textsc{phonopy} package~\cite{phonopy}. The following supercell sizes with respect to their respective primitive cells were employed: 2$\times$2$\times$2, 2$\times$1$\times$1, 2$\times$1$\times$2, and 3$\times$3$\times$3 for the FiE, AFE$_{80}$, AFE$_{40}$, and FE, phases, respectively. We consider the non-analytical contribution to the phonons in all the calculations~\cite{gonze97}. The temperature dependent free energies were derived within the harmonic approximation, as implemented also in \textsc{phonopy}~\cite{phonopy}. The phononic integrals are performed in the BZ over $\bb{q}$-meshes twice as dense in each direction as the electronic ones (10$\times$10$\times$10, 16$\times$8$\times$6, 16$\times$8$\times$12, and 16$\times$16$\times$16 for the FiE, AFE$_{80}$, AFE$_{40}$, and  FE phases, respectively). We checked that these meshes yield zero point energies converged within 0.01~meV by comparing with meshes twice as dense along each direction. In the AFE$_{40}$ phase, where an unstable phonon branch appears near the $Z$ point, the unstable modes (which correspond to only 0.05\% of all the modes) are excluded from the summations. The cubic perovskite features several strong instabilities spanning the whole Brillouin zone (see Supplementary Figure 1), and hence excluding the instabilities from the summations would result in very large errors, so we cannot compute the free energy for this state.

The zero-point energy, $E_\mathrm{ZPE}$, is given by $E_\mathrm{ZPE}=\frac{1}{2}\sum_{\bb{q}\nu}\hbar\omega_{\bb{q}\nu}$, where $\omega_{\bb{q}\nu}$ is the phonon branch with wavevector $\bb{q}$ and index $\nu$, and $\hbar$ is the reduced Planck constant. The free energy is computed as
\begin{equation}
		 F=E_\mathrm{KS}+E_\mathrm{vib}-TS
 \end{equation}
where $E_\mathrm{KS}$ is the Kohn-Sham energy as obtained directly from the DFT calculation, and $E_\mathrm{vib}$ is the vibrational energy given by
 \begin{equation}
		E_\mathrm{vib}=\sum_{\bb{q}\nu}\left[ \frac{1}{2} + n_{\bb{q},\nu} \right] \hbar \omega_{\bb{q},\nu} = E_\mathrm{ZPE}+\sum_{\bb{q}\nu} \hbar \omega_{\bb{q},\nu} n_{\bb{q},\nu},
 \end{equation}
 where $k_B$ is the Boltzmann constant and $n_{\bb{q},\nu}$ the number of phonons with wavevector $\bb{q}$ and branch index $\nu$, $n_{\bb{q},\nu}=\left({\mathrm{exp}(\hbar\omega_{\bb{q}\nu}/k_BT)-1}\right)^{-1}$.
 The entropy $S$ reads
 \begin{equation}
	 \begin{split}
S= -k_B \sum_{\bb{q}\nu} \mathrm{ln}\left[1-\mathrm{exp}(-\hbar\omega_{\bb{q}\nu}/k_BT)\right] 
		 +\frac{1}{T}\sum_{\bb{q}\nu} \hbar\omega_{\bb{q}\nu}n_{\bb{q},\nu}
	 \end{split}
 \end{equation}
 so we can write the free energy as
 \begin{equation}
		 F=E_\mathrm{KS}+E_\mathrm{ZPE}+k_BT \sum_{\bb{q}\nu} \mathrm{ln}\left[1-\mathrm{exp}(-\hbar\omega_{\bb{q}\nu}/k_BT)\right]
 \end{equation}
 In the free energy calculations shown in Figure~\ref{fig_freeenergy} the phonon modes $\omega_{\bb{q}\nu}$ are assumed to be independent of volume and temperature. We considered the effect of the thermal expansion on the FiE and AFE$_{40}$ phases by computing the Gibbs free energy ($G$) of both phases at zero pressure within the quasi-harmonic approximation. The Gibbs free energy is given by
 \begin{equation}
	 \begin{split}
	 G(T,P)=\min_V\left( F(T,V)+P V \right) 
	 \end{split}
 \end{equation}
 where $P = -(\partial F / \partial V)_{T}$ is the pressure and $V$ is the volume. To this end, we computed the Kohn-Sham and vibrational energies of the two phases under $0.5\%$ and $1.0\%$ homogeneous expansive strain. The Kohn-Sham energies are interpolated for an arbitrary volume using a Birch-Murnaghan equation of state~\cite{birch,murnaghan}, while the vibrational energies are interpolated using second order polynomials of the volume at each temperature, scanning the temperature in 1~K intervals in the 0-500~K range. Note that even at $T=0$, at the volume that minimizes the Kohn-Sham energy, the zero-point energy induces a positive pressure. Therefore, the zero-pressure Gibbs free energy shown in Figure~\ref{fig_freeenergy}{\bf d} and the free energy in dashed lines in the same figure do not coincide at $T=0$.

{\bf Structure visualization.}
We used the \textsc{vesta} visualization package~\cite{VESTA} to prepare some of the figures.

\vspace{5mm}{\bf Data availability}
All the relevant data are available from the authors upon reasonable request.

\vspace{5mm}{\bf Code availability}
The first-principles calculations are carried out using the open source package \textsc{vasp}, which is a proprietary software. The space groups of the crystal structures and phonon dispersions are obtained using the open source package \textsc{phonopy}, which is released under the BSD-3-Clause License (https://github.com/atztogo/phonopy). The \textsc{isotropy} suite is an open source package freely available at https://stokes.byu.edu/iso/isodistort.php. The visualisation software \textsc{vesta} is distributed free of charge for scientific users under the \textsc{vesta} license (https://jp-minerals.org/vesta/en/download.html).

\vspace{5mm}{\bf Acknowledgements}
{H}.{A}. and {J}.\'{I} acknowledge funding by the Luxembourg National Research Fund through the project INTER/ANR/16/11562984/EXPAND/Kreisel. {C}.{C}. acknowledges support from the Spanish Ministry of Science, Innovation and Universities under the ``Ram\'on y Cajal'' fellowship RYC2018-024947-I. {M}.{S}. acknowledges the support of Ministerio de Econom\'ia, Industria y Competitividad (MINECO-Spain) through Grant No. PID2019-108573GB-C22 and Severo Ochoa FUNFUTURE centre of excellence (CEX2019-000917-S); of Generalitat de Catalunya (Grant No. 2017 SGR1506); and of the European Research Council (ERC) under the European Union's Horizon 2020 research and innovation program (Grant Agreement No. 724529). 

\vspace{5mm}{\bf Author contributions}
H.A. performed the first-principles study, aided by C.C. for the free-energy calculations, and supervised by J.\'I. M.S. proposed the FiE phase and conceived the work together with H.A. and J.\'I. All authors contributed to the discussion and analysis of the results. The manuscript was written by H.A. and J. \'I., with contributions from C.C. and M.S.

\vspace{5mm}{\bf Competing interests}
The authors declare that they have no competing interests.

\bibliographystyle{naturemag_nourl}

{\bf Figure legends.}

Figure 1. {\bf Crystal structure of the AFE$_{40}$ and FiE phases of PbZrO$_3$.} 
	Top ({\bf a},{\bf c}) and side ({\bf b},{\bf d}) views of the  AFE$_{40}$ (left panels) and FiE states (right panels) are displayed. The primitive cells are marked with a black line. The Zr and O atoms are shown in green and black, respectively. Pb atoms are coloured in red and blue for positive and negative displacements, respectively. The oxygen octahedra are coloured in green and yellow alternatively (according to their displacements given by the $R$-point instability) as a guide to the eye.

Figure 2. {\bf Schematic Pb displacements in the AFE$_{40}$ and FiE phases of PbZrO$_{3}$.} Top view of the in-plane displacements of the lead cations in the AFE$_{40}$ phase ({\bf a}), and in the FiE phase ({\bf b}). Unit cells marked in black; lead cations marked in red and blue. For the FiE phase a smaller primitive cell (of only 30 atoms) exists, for which the in-plane projection is shown with a dotted line in  {\bf b} (see also Fig.~\ref{fig_crystal}). As explained in the text, in {\bf b} the two red cations in the primitive cell move by $+$0.16~\AA\ and $+$0.23~\AA, respectively, while the blue cation moves by $-$0.21~\AA.

Figure 3. {\bf Kohn-Sham energies as a function of volume per formula unit for the PbZrO$_3$ polymorphs at different pressures and with different exchange-correlation functionals.} LDA results for 0, $-$4 and $-$8~GPa are displayed (bottom). PBEsol and SCAN results shown for $+$4, 0 and $-$4~GPa (centre, solid and dashed lines, respectively). PBE results for $+$8,	$+$4 and 0~GPa shown (top). The energies are shown relative to the energy of the cubic phase for each pressure. Blue circles, red empty triangles, green filled triangles, and black squares correspond to the FiE, AFE$_{40}$,	AFE$_{80}$, and FE phases, respectively. The results for SCAN are shifted by $+$20~meV per f.u. for clarity.
	
Figure 4. {\bf Phonon band structure and phonon density of states of the studied PbZrO$_{3}$ phases computed with PBEsol.} The results for the FiE , AFE$_{80}$, AFE$_{40}$, and FE states are shown in panels {\bf a}, {\bf b}, {\bf c}, and {\bf d}, respectively.
	
Figure 5. {\bf Free energies of the studied PbZrO$_{3}$ phases as a function of temperature.} Free energy differences to the AFE$_{40}$ phase (red) as a function of temperature for the FiE (blue), AFE$_{80}$ (green), and FE (black) phases of PbZrO$_3$ are shown in  panels {\bf a}, {\bf b}, {\bf c} and {\bf d} for PBEsol, LDA, PBE, and SCAN, respectively. The inset in {\bf a} shows the Gibbs free energy difference to the AFE$_{40}$ phase at zero pressure for the FiE state using PBEsol. The dashed line in the inset corresponds to the free energy of the FiE without considering the thermal expansion.

\end{document}